\documentclass[journal]{IEEEtran}
\usepackage{cite}
\usepackage{xcolor}
\usepackage{epsfig}
\usepackage{graphicx}
\usepackage{epstopdf}
\usepackage{balance}
\usepackage{url}
\usepackage[caption=false]{subfig}
\usepackage{eurosym}
\usepackage{threeparttable}
\usepackage[cmex10]{amsmath}
\usepackage{array}
\usepackage{booktabs}
\usepackage{tablefootnote}
\usepackage{enumerate}
\usepackage{multirow}
\usepackage{threeparttable}

 \newcommand{\trnote}[1]{\TPTrlap{\tnote{#1}}}

\ifCLASSINFOpdf
\else
\fi
\begin{document}
%
\title{A sub-mW IoT-endnode for always-on visual monitoring and smart triggering}
%
%
%

\author{Manuele~Rusci,~
        Davide~Rossi,~
        Elisabetta~Farella,~
        and~Luca~Benini

\thanks{M. Rusci and E. Farella are with Fondazione Bruno Kessler, Trento, Italy. M.Rusci, D. Rossi and L. Benini are with University of Bologna, Italy. L. Benini is also with ETH Zurich, Switzerland.}
}

\maketitle

\begin{abstract}
This work presents a fully-programmable Internet of Things (IoT) visual sensing node that targets sub-mW power consumption in always-on monitoring scenarios. 
The system features a spatial-contrast $128\mathrm{x}64$ binary pixel imager with focal-plane processing. 
The sensor, when working at its lowest power mode ($10\mu W$ at 10 fps), provides as output
the number of changed pixels. 
Based on this information, a dedicated camera interface, implemented on a low-power FPGA, wakes up an ultra-low-power parallel processing unit to extract context-aware visual information.
We evaluate the smart sensor on three always-on visual triggering application scenarios. Triggering accuracy comparable to RGB image sensors is achieved at nominal lighting conditions, while consuming an average power between $193\mu W$ and $277\mu W$, depending on context activity. The digital sub-system is extremely flexible, thanks to a fully-programmable digital signal processing engine, but still achieves 19x lower power consumption compared to MCU-based cameras with significantly lower on-board computing capabilities. 
\end{abstract}

%
\IEEEpeerreviewmaketitle


\section{Introduction}
\label{intro}
The Internet of Things (IoT) paradigm envisions a wide infrastructure network of "things" that forms a pervasive computing environment \cite{Chen2014}. Many interconnected objects will be able to sense physical phenomena and exchange data, information and knowledge through the network, to leverage the user experience of the surrounding environment.
Energy efficiency is a key aspect for these IoT battery-powered devices, which feature sensing, communication and processing capabilities. 
Typically, the communication sub-system is dominant in terms of system power consumption because of the high amount of raw data to be transmitted \cite{deGroot2015}.
Near sensor processing allows to overcome this limitation by locally analyzing data coming from the sensors. Following this approach, the data transmission operation involves compressed information or even a simple alarm bit that is triggered when an interesting event has been detected. To accomplish the in-sensor filtering tasks, a smart sensor couples a digital processor to one or more sensing units. 
Also, by exploiting a parallel data processing, the event detection process presents a reduced latency with respect to a sequential data analysis \cite{Mayer2015}.
From an energy prospective, several system level power management strategies, such as duty-cycling and dynamic voltage and frequency scaling, can be implemented to maximize the system battery life \cite{Dargie2012, casamassima2013}.
Unfortunately, duty cycling and aggressive frequency scaling reduce system responsiveness to sporadic events. This goes against the always-on requirement of many IoT applications, where low sampling rates or wakeup delays directly impact application quality metrics. 

\begin{figure}[]
	\centering
  	\includegraphics[width=1\linewidth]{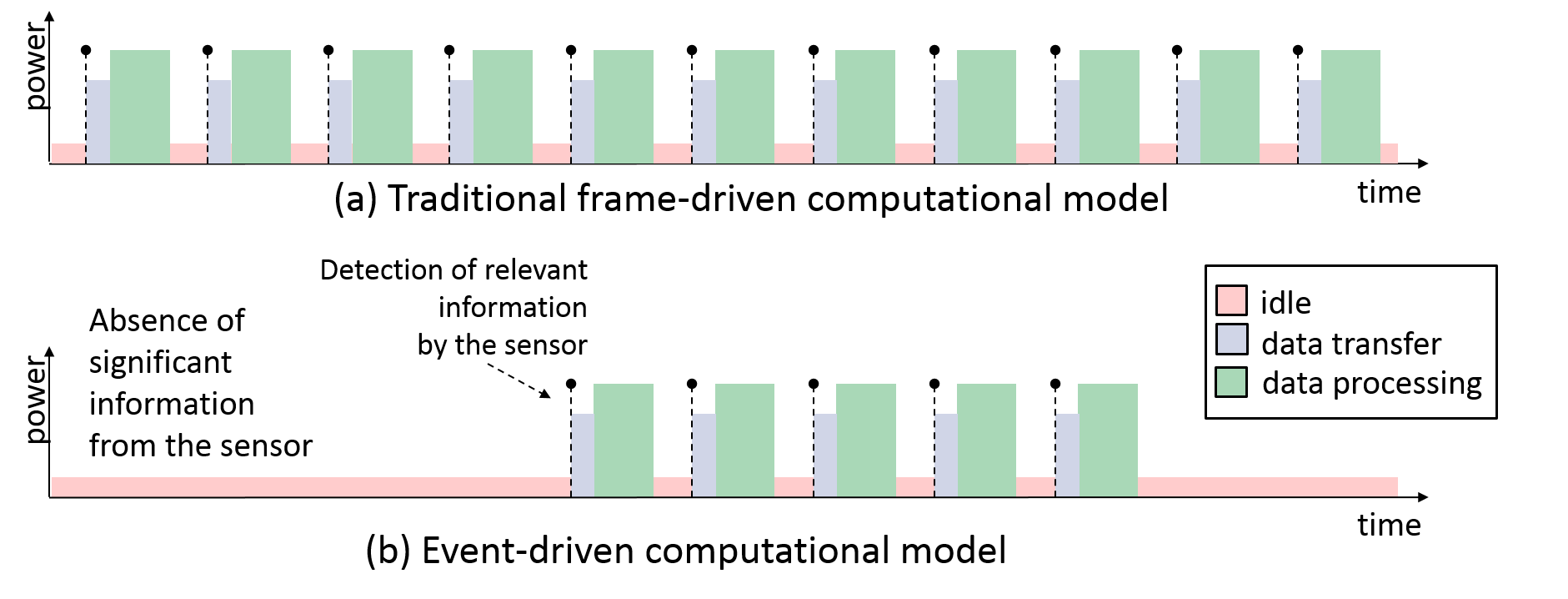}
	\caption{Energy-efficiency comparison between frame-driven and event-driven computational frameworks.}
	\label{fig:EventDriven}       
\end{figure}

In the context of visual sensor network, 
an energy-efficient computational framework consists of locally analyzing the huge amount of raw visual data before the transmission of the extracted information \cite{Bondi2015}. A flexible and low-power hardware platform is required to implement such paradigm. 
In recent years, several smart cameras have been proposed to develop intelligent and autonomous systems \cite{Tavli2012}. Typically, such devices feature a power consumption of hundreds of \textit{mW}, because of the high computational and bandwidth requirements \cite{Chen2013,CMUCAM,Bondi2015}. 
Power-optimized solutions rely on vision chips that integrate focal-plane processing circuits, which enable a first stage of visual processing in a distributed and efficient way at the sensor-level \cite{Fernandez2016}. 
To extract digital signatures from the visual signal, sensor data are then transferred to the digital processing unit for post-processing operation.
A most widespread frame-based computation paradigm is based on frame-by-frame data analysis \cite{Ko2007}.
Although this mainstream approach is effective, it is power-inefficient because it requires to periodically wake-up the digital processing unit, even if no interesting elements appear on the camera field of view. In this paper we claim and demonstrate that, from an energy perspective, an event-driven computational model represents a more efficient framework when ultra-low-power always-on functionality is needed. According to such model, which is illustrated in Fig. \ref{fig:EventDriven}, the sensing unit itself selectively activates the digital processor only if some relevant external events are detected by the sensor. 
As a consequence, the average power consumption is reduced with respect to the traditional frame-driven approach (also referred as periodic-polling) thanks to the much more sporadic need for data transfer and processing operations. Potentially, the average system power consumption can reach the idle power in case of rare or very infrequent external events.

In this work we present the implementation of a fully-programmable smart camera IoT-endnode based on the concept originally introduced in \cite{Rusci2016}. The visual node relies on an ultra-low-power vision sensor, which exploits analog processing circuits to extract pixel-wise contrast information. A parallel digital processing unit (PULP\footnote{Parallel Ultra Low Power - http://www.pulp-platform.org/ [accessed on: 2016-07-30]} \cite{PULP}) is coupled to the imager to enable an efficient and flexible data processing. The sensor features a low-power mode, named \textit{Idle}, where it delivers only the number of asserted binary pixels. In \textit{Active} mode, instead, a stream of coordinates related to the asserted pixels is readout. The main contributions of this paper are:
\begin{itemize}
\item The design of a power-optimized camera interface and its implementation on a low-power FPGA.
\item The implementation of a context-aware system-level power management strategy, which is evaluated within three different visual triggering application scenarios.
\end{itemize}

According to the event-driven framework, the camera interface is responsible for triggering the processor wake-up event when the sensor, normally working in \textit{Idle} mode, detects a relevant information.
With respect to traditional systems, the always-on functionality metrics are not influenced by processor shutdowns and duty cycle. Here, the processing activity rate depends on the content of the context. Potentially, the system is constantly kept in Idle state when no interesting information are detected by the imager.
In the paper, we evaluate the triggering process by using our contrast-based vision chip and a comparison with respect to traditional approaches conducted over usual camera data (i.e. RGB images). We conclude that the employed approach does not lead to performance degradation over the considered application case-study unless the scene illumination is considerably reduced.

Within the considered monitoring applications, the presented system can reach an overall power consumption of $277\mu W$ thanks to the implemented power-management strategies and by selectively activating the digital processing unit. Such consumption is an order of magnitude lower than other reported smart camera systems. 
Moreover, the fully-programmable processor allows a higher flexibility than customized hardware processing circuits.

The reminder of this paper is organized as follows. Section \ref{sec:RelWork} summarizes the related work. The system components are described in Section \ref{sec:SysArc} and in Section \ref{sec:cameraIF} the implemented camera interface is presented. Section \ref{sec:Prot} presents the adopted power management strategy and Section \ref{sec:App} describes the considered visual applications and the employed triggering process approach. The experimental evaluation, both in term of detection accuracy and power consumption, is reported in Section \ref{sec:Eval} while Section \ref{sec:concl} concludes the paper.

\section{Related Work}
\label{sec:RelWork}

Low power and low latency are key requirements for IoT sensing platforms\cite{Chen2014}. 
Especially in the context of visual sensor nodes, power-efficiency is a great concern because of the remarkable power costs of the components and the high-bandwidth and computational requirements of visual applications \cite{Abas2014}. Commercially available products, such as CMUcam5 \cite{CMUCAM} and the Blackfin Low Power Imaging Platform (BLIP) \cite{BLIP}, show a power consumption of several hundreds of mW on typical application workload. A survey of existing visual nodes is reported in \cite{Tavli2012}.
Among them, CITRIC \cite{Chen2013}, which includes a PXA2700 32-bit processor, is used in \cite{Casares2011} to perform object detection and tracking with a power budget of 751mW. To save energy, a reduced amount of data is transferred from the sensor to the MCU, thanks to hardware-level down-sampling and cropping.
Mesheye \cite{Hengstler2007} employs a heterogeneous camera architecture for object detection and tracking. A high-resolution camera acquisition is triggered when an object is detected by means of a low-resolution stereo camera.
Cyclops \cite{Rahimi2005} features a CIF (352x288) resolution camera and an ATMEL ATmega128L 8-bit processor. An additional CPLD acts as frame-grabber when enabled by the MCU. A power consumption of 33mW is reported when the node is employed for triggering wake-up signals to upper layer network systems \cite{Kulkarni2005}.
All these reported systems feature traditional power-hungry imagers and digital platforms, which are not fully optimized for low-power consumption.

To increase the power-efficiency of camera-based architectures, focal plane processing have been investigated \cite{Fernandez2016}. According to this approach, pixel-wise mixed-signal processing circuits are integrated on the sensor die to enable a first stage of processing acceleration in a distributed and efficient way \cite{Kim2013,Park2014,Choi2014}. 
As a result, the sensor dispatches out pre-analyzed data, reducing the post-processing workload of the digital processing unit. Wi-FLIP \cite{Fernandez2011} exploits the inherent parallelism of early-vision tasks thanks to FLIP-Q \cite{Fernandez2012}, a 176x144 pixels imager that embeds focal-plane analog functionalities. Within this smart camera system, a PXA271 processor is coupled to the sensor, leading to a power consumption of 126mW. 
Despite the available computational power and the proven functionalities, the power consumption of this system still exceeds the targeted consumption for always-on IoT applications due to the not-optimized processing platform. 
A similar architecture to the one reported in this work is described in \cite{Carey2013}. A low-power smart camera contains a vision chip with a large array of processing elements and a low-power FPGA, which is placed between a digital MCU and the sensor. The MCU is put in deep-sleep mode during the light integration period. The FPGA, which is fed by a 32kHz RTC, eventually triggers the MCU's wake-up for image readout and processing. By exploiting the sleep power mode and duty cycling, which results to be limited by the power supply and oscillator start-up times, authors report an average power consumption of 5.5mW within a surveillance application. 
Unlike this smart camera, our systems is continuously kept in idle mode unless a relevant information is detected by the always-on sensor. Such event-driven computational model allows to save the additional energy required to wake-up the processor when no interesting event is captured within the camera field-of-view. This strategy has been employed by multimodal visual sensors, such as \cite{Magno2013}, where a PIR sensor is used to trigger the smart camera activation. With respect to this, our system implements a wake-up process based on the visual data itself, which contains much richer information compared to a monodimensional signal, and can be eventually applied to a wider range of applications. 

A people counter application has been implemented and presented in \cite{Gasparini2011} by using the same vision chip as in our architecture. 
The presented system relies on a flash-based FPGA for data processing.
Despite the relatively low-power consumption of 4.2mW, the system lacks in flexibility due to the hardwired processing function. Moreover, authors address a single person counter while our camera node allows to detect and track multiple moving objects thanks to the software fully-programmable architecture \cite{RusciISC2}. 
The event driven visual processing concept was firstly introduced in \cite{Rusci2016}. In the current work we describe a complete system design and implementation and the evaluation of the adopted power management strategies on three different monitoring application scenarios. With respect to the previous work, a low-power FPGA is employed to implement a power-optimized camera interface, which is in charge to wake-up the processing unit depending on the context-activity.

Relevant work on event-driven visual sensing has been published in the neuromorphic engineering community \cite{Posch2014}. Asynchronous event-based cameras \cite{Delbruck2008,Lichtsteiner2008,Lenero} have been employed for robotics \cite{Delbruck2015} and monitoring applications \cite{Litzenberger2006}. Such sensors asynchronously readout pre-processed events when intensity or contrast changes are detected at pixel-level.  
Embedded low-power implementations of event-based visual platforms have been presented in \cite{Litzenberger2006} and \cite{Muller2011}. In the latter, a power consumption of $23mW$ is measured. However, at the best of our knowledge, power management strategies have not been reported yet to address power-related issues. 
With respect to this kind of asynchronous systems, our proposed architecture employs a frame-driven timing for data readout. This feature allows the presented system to exploit the lowest power mode between successive readout on every components, resulting in a remarkable system power saving.

\begin{figure*}[t]
	\centering
  	\includegraphics[width=\textwidth]{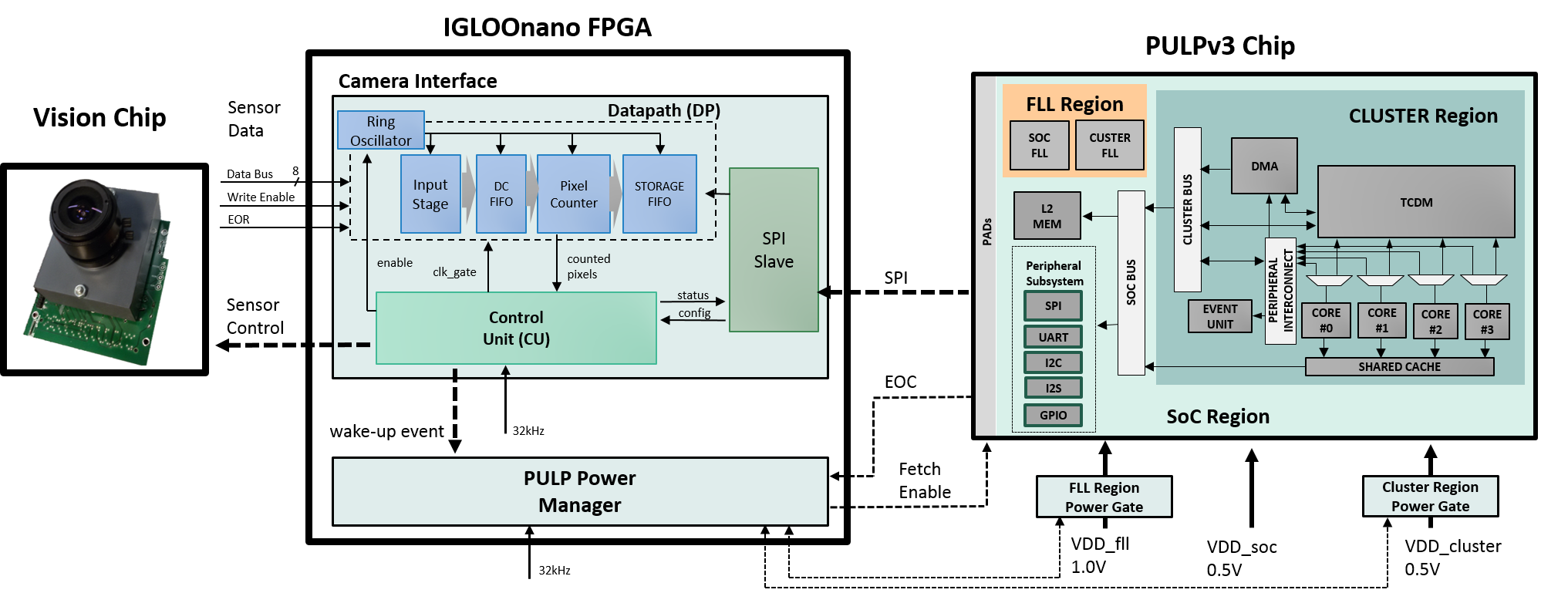}
	\caption{Architecture of the camera interface when implemented as external component.}
	\label{fig:archi}       
\end{figure*}
\section{System Architecture}
\label{sec:SysArc}
In this section, an overview of the system components is provided. The vision chip and the processing unit are described below.

\subsection{Vision Chip}
\label{sec:grain}
The $128\mathrm{x}64$ imager \cite{Gottardi2009} embeds pixel-wise analog circuits to compute the spatial contrast among neighbouring pixels. The information is binarized, thanks to analog comparators, and stored through an internal pixel memory. A frame difference operation between successive frames allows to individuate binary pixels corresponding to moving objects. 

The vision chip features a native interface, which consists of control and data IO pins. To operate a continuous focal-plane processing, a periodic control sequence is provided in input, as detailed in \cite{Gottardi2009}. Regarding the output of data, the sensor features two different readout modes. In \textit{Idle} mode, the asserted pixels (i.e. pixels with value +1 or -1 after the binary frame difference) are internally counted and only the counter value is shown as output. In \textit{Active} mode, instead, the asserted pixels are dispatched out according to a raster-scan readout process. 
The external sensor interface samples the asserted pixels, which are dispatched through a 8-bit Data Bus, on the rising edge of the Write-Enable output pin. 
Every pixel information is composed by 7 bits representing the column coordinate (\textit{y-}coordinate) and 1 bit for the sign.
A End-Of-Row (\textit{EOR}) output pin indicates the increase of the row index during the readout process. Therefore, the pixel \textit{x-}coordinate is retrieved by the sensor interface by taking into account the number of EOR pulses up to the pixel reception.
The data streaming is considered terminated when the \textit{EOR} signal have pulsed for 64 times (the number of rows). The peak data rate reaches $80Mpixel/sec$ within the readout process. 

As also reported in \cite{Gottardi2009}, the sensor power consumption depends on the readout mode. In \textit{Idle}, the power consumption is mainly due to the sensor internals. A power consumption of 10$\mu W$ has been measured at 10fps and 25\% pixel activity. In \textit{Active} mode an additional energy consumption is incurred due to the increased pad activity, which is proportional to the number of asserted pixels and to the frame rate.

\subsection{PULP}
		\label{sec:pulp}	
PULP (Parallel Ultra Low Power) is a multicore processing system targeting high-energy efficiency. 
The PULPv3 System on Chip (SoC) 
\cite{PULP} includes a 4-core cluster and several IO peripherals. A 28nm FD-SOI chip prototype has been designed and fabricated. 
The SoC architecture is depicted in Fig. \ref{fig:archi}. A cluster region includes the processing cores, along with a 48kBytes Tighly Coupled Data Memory (TCDM), acting as software-managed L1 memory, and a Direct Memory Access (DMA) engine to handle data transfer with L2 memory. Among the cluster peripherals, the event unit autonomously handles the clock-gating of individual cores in idle state (e.g. if waiting at a synchronization barrier).

The off-cluster region, also named SoC region, contains a 64kBytes L2 memory and the peripheral subsystem, which includes several IO interfaces, such as SPI, UART and I2C. The cluster and the SoC regions exploit different power domains to enable a flexible dynamic voltage scaling. Also, the two regions feature different clock domains thanks to dedicated Frequency Locked Loops (FLLs), which are placed in the so-called FLL region. This latter is powered at 1V. The FLLs are fed by a 32kHz external clock oscillator and allow a fine-grain tuning of the SoC and cluster clock frequencies. 

PULP is a fully programmable platform leveraging C and OpenMP for software programming. This leads to a complete flexibility for application development and eases the design of the signal-processing flow.

\section{Camera Interface Design and Implementation}
\label{sec:cameraIF}
From a system-level point of view, the camera interface is in charge of:
\begin{enumerate}
\item handling the communication between the vision chip and the PULP platform, and
\item autonomously managing the sensing operation and the power modes transitions of the imager without the processor intervention.
\end{enumerate}


Figure \ref{fig:archi} details the camera interface block, which is placed between the vision chip and the processor.
The processing unit sets up the camera parameters and reads out the sensor data through the SPI slave port. For instance, the frame-rate and the exposure time have to be defined by means of SPI write transactions.

The camera interface architecture contains a Control Unit (CU) and a DataPath (DP) modules. The first one handles the sensor control signals depending on the current readout mode and the specified parameters. A detailed description of the sensor signals and the associated timing is described in \cite{Gottardi2009}. The CU also manages the transition scheme between \textit{Idle} and \textit{Active} modes based on a user-defined threshold, as also proposed by \cite{Gasparini2011}. 
When the sensor is \textit{Idle} mode, the camera interface readouts the counter value of asserted pixels through the sensor data bus. If the value overcomes the threshold, the CU switches the imager to Active mode by generating the proper timing signals. 
The Datapath serves for gathering and storing data coming from the sensor in Active mode. Such submodule firstly converts visual data from the native protocol to a \textit{(x,y)} format, where x- and y- stand for the camera plane coordinates, and then counts and store them into an internal buffer (Storage memory). 
Once the readout process is completed, the DP is disabled and all the received data are available in the Storage memory. Following this, the CU triggers a wake-up event to the processor to request data transfer and processing.

The camera interface submodules feature opposite requirement in terms of clock speed. Indeed, the CU is fed by low-speed free-running clock to allow the continuous sensor timing control. A reference clock of 32 kHz is employed for this scope. The DP clock, instead, needs to be sufficiently high to sustain the sensor peak output rate of $80Mpixel/sec$ during the short readout period. To save power, this high-speed clock, which is generated by a ring-oscillator, can be disabled for the rest of the time of the frame period. With respect to the circuit described in \cite{Gasparini2011}, the presented architecture includes separated clock sources for the two clock domains. Therefore, the ring-oscillator can be turned-on by the always-on Control Unit only during the readout period when the sensor is in \textit{Active} state. Considering a readout period of about $300\mu sec$ and a frame period of $\mathrm{100msec}$ (corresponding to $\mathrm{10fps}$), a power saving of 39x is achieved by disabling the ring-oscillator for the majority of the frame time with respect to keep both the clock drivers continuously active.

\begin{figure*}[t]
	\centering
  	\includegraphics[width=0.8\textwidth]{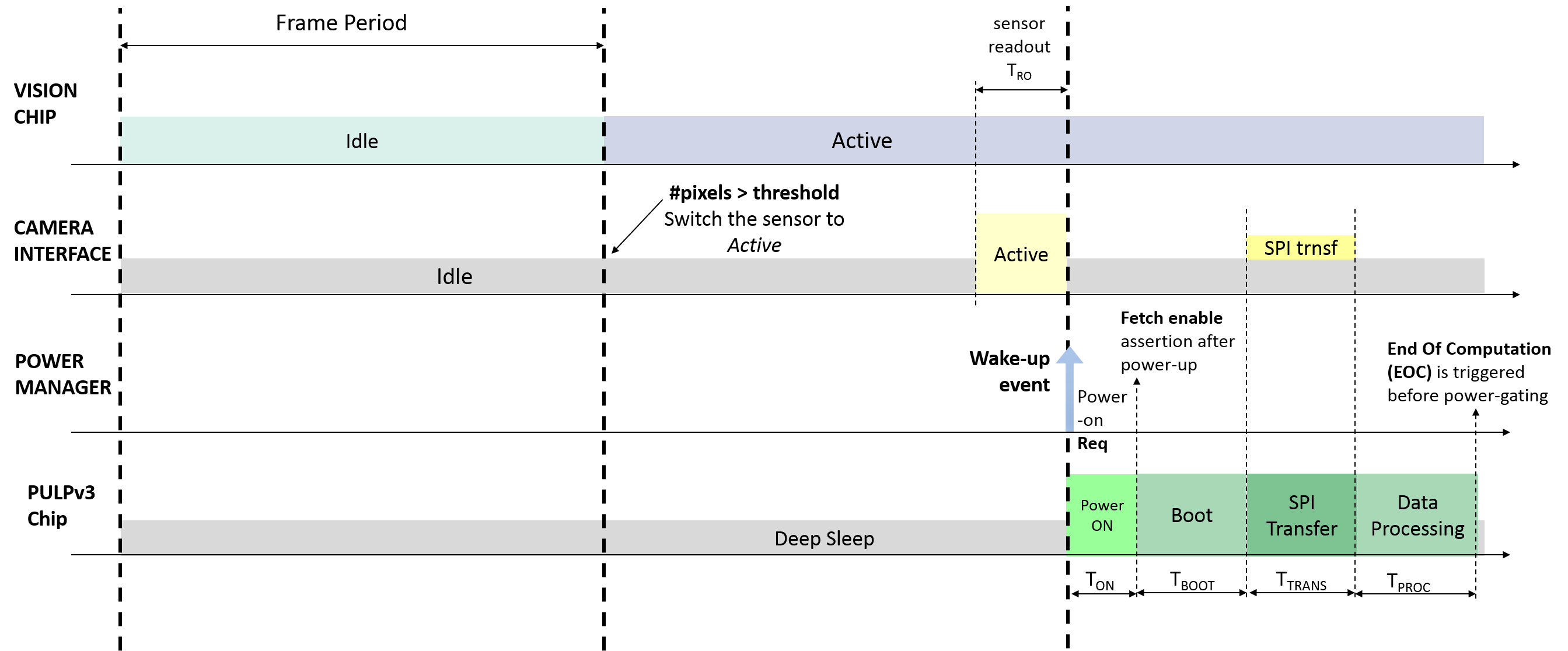}
	\caption{Power management strategy within the smart camera system.}
	\label{fig:state}       
\end{figure*}

\begin{table}[b]
\caption{FPGA Resource Utilization}	\label{tab:resource}
    \centering

\begin{threeparttable}
\centering
\resizebox{1\columnwidth}{!}{
    \begin{tabular}{ | l | l | l |  l |}	
    \hline
	Resource   & Used & Total & Percentage	\\
	\hline \hline
	VersaTiles \trnote{a} & 3758 & 6144 & 61.17\% \\
	RAM 4,608-Bit Blocks & 8 & 8 & 100\% \\
	VersaNet Globals \trnote{b} & 6 & 18 & 33.33\% \\
	\hline
    	\end{tabular}
}
\begin{tablenotes}
\item [a] {\scriptsize{Equivalent to a three-input lookup table (LUT)
or a D-flip-flop/latch with enable}}
\item [b] {\scriptsize{Global clock distribution network}}
\end{tablenotes}
\end{threeparttable}
\end{table}

The camera interface has been implemented on a low-power flash-based FPGA Microsemi's IGLOOnano AGLN250V2 \cite{IGLOO}.
An external 32 kHz clock oscillator drives the Control Unit while the high-speed clock is internally generated through the ring-oscillator.
When the FPGA core works at 1.2V, which is the lowest nominal supply voltage, critical timing issues are reported if the high speed clock is at 80Mhz, which corresponds to the peak sensor output rate.
To overcome this bottleneck, an input register within the datapath gathers 4 incoming pixels before pushing them into the DC FIFO. 
Thus, to relax the timing constraints, the ring-oscillator frequency is set to 25MHz, which results to be sufficiently high to handle the transfer of 4-pixels packets to the Storage Memory. Tab. \ref{tab:resource} reports the FPGA resource utilization.
The internal memory buffer (Storage Memory) has been sized to collect a maximum amount of 1024 pixels, corresponding to 12.5\% of the theoretical maximum amount, which is sufficient to store the amount of pixels in most common typical monitoring application scenarios \cite{Rusci2016}. However, if a failure occurs causing the loss of sensed data, a visual tracking-based post-processing approach, as the one employed within the presented monitoring applications, is sufficiently robust to recover the missed information from the prior extracted data.

\begin{table}[]
\caption{Power consumption of the system components}	\label{tab:power}

\centering
\resizebox{1\columnwidth}{!}{
	\begin{tabular}{ |l|l|l| l|p{3cm}|}
	\hline
	 \multicolumn{2}{ |l| }{Component}  & Fully-Active Power  & Idle Power & Note \\ \hline \hline
	\multicolumn{2}{ |l| }{Vision Chip}
 		&  $20\mu W$ &  $10\mu W$ 
 		& Measured at 10fps and 25\% activity \cite{Gottardi2009}\\
	\hline
	 \multicolumn{2}{ |l| }{FPGA Camera	IF} 
 			&  $3mW$ & $68\mu W$ &  Additional $456\mu W$ during SPI transfer at 5 MHz \\
	\hline
	\multirow{3}{*}{PULPv3 Chip} 
 		& Cluster Region &  $946\mu W$ & - & At 0.5V, 30MHz \\  \cline{2-5}
  		& SoC Region & $313\mu W$ & $99\mu W$ & At 0.5V, 30MHz \\  \cline{2-5}
 		& FLL Region & $3.2mW$ & - & At 1V
 		 \\  
	\hline
	\end{tabular}
}
\end{table}

\section{System Power Management Strategy}
\label{sec:Prot}
The vision chip, once coupled to the camera interface, enables a context-aware system level power management strategy. 
According to this strategy, here also referred as Event-Driven, 
the device exploits information from the context to select the appropriate power mode.
In this case, the system is fully activated only when a relevant information is detected by the sensor. During the rest of the time, instead, the components are kept in a Idle state to consume a minimal power. Still, the continuous sensing operation is guaranteed thanks to the imager and the camera interface.

\begin{figure}[b]
	\centering
  	\includegraphics[width=0.6\linewidth]{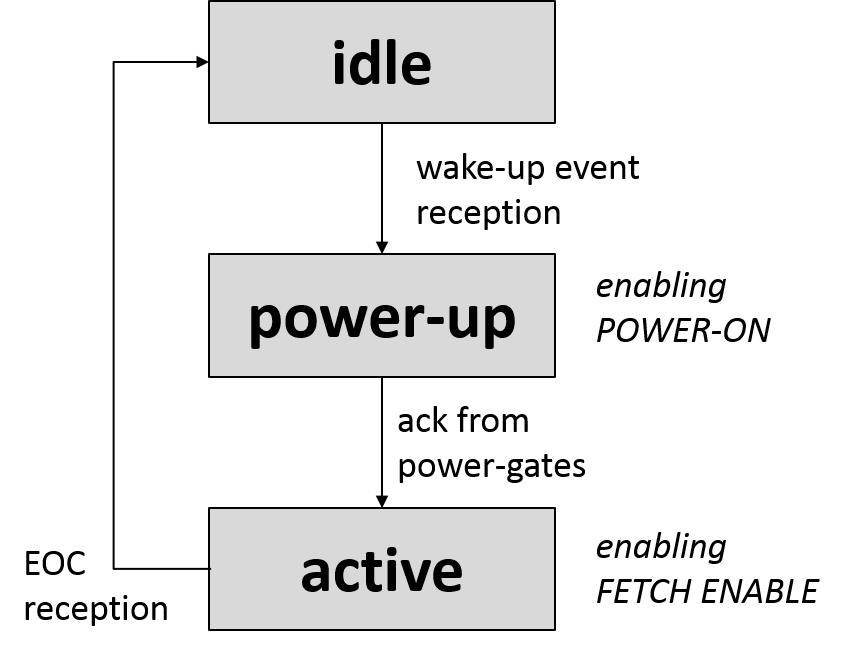}
	\caption{PULP External Power Manager state machine}
	\label{fig:fsm}       
\end{figure}

Tab. \ref{tab:power} reports the measured components power consumption, either in Fully-Active and in Idle states. When operating at the peak capacity, the camera interface consumption is dominated by the dynamic power associated to the high-speed clock domain. When the ring oscillator is disabled, the FPGA power is due to the leakage and the 32kHz clock activity, resulting in an extremely lower value of $68\mu W$. Therefore, to save power when the sensor is in \textit{Active} state, the ring-oscillator is activated only during the short period of data readout (Tab. \ref{tab:timing}). Acting such duty cycle technique, an average power of $77\mu W$ is achieved within the frame period. In addition to this, a power contribution of $456\mu W$ is accounted during the SPI data transfer. In Idle state, instead, the FPGA power consumption is limited to $68\mu W$.
Tab \ref{tab:power} also reports the measured consumptions associated to each PULP power domain. Both the cluster and the SoC regions are powered at 0.5V while the FLL region requires a higher supply voltage of 1V. The Idle state of the processor refers to the implemented deep-sleep mode. Within such state, the FLL and the Cluster regions are power-gated, while the SoC region remains continuously powered to guarantee data and executable code retention on the L2 memory.  This is essential for a fast start-up process since loading the executable (45kBytes) from an external non-volatile memory requires 1.7msec through a quad-SPI interface at 50 MHz\cite{FLASH}. Moreover, even assessing the increased latency as a non-critical issue, the energy consumption associated to the read transfer operation from the external flash memory results to be 7x higher than the energy consumed within the frame period due to the leakage of the always-on SoC region.
An external controller is implemented on the FPGA, along with the camera interface, to manage the transitions between Active and Deep Sleep mode of the PULP processor through discrete on-board power gate components. The internal state machine of the PULP power manager is illustrated in Fig. \ref{fig:fsm}. After the reception of the wake-up event from the camera interface, the power manager is in charge of powering-on the PULP processor by controlling the power-gates. The fetch-enable signal is then asserted after the reception of the acknowledge from the power-gates components. Finally, the controller goes back in Idle state once the End-Of-Computation (EOC) signal is asserted by the PULP processor.
Fig. \ref{fig:archi} illustrates the implemented system scheme to realize the addressed power management strategy.

Tab. \ref{tab:timing} lists the start-up timing delays of the PULP platform when powered-on. Specifically, $T_{ON}$ refers to the power-on time of the supply voltages and the FLL lock time when in closed-loop configuration. $T_{BOOT}$ indicates the boot-up time of the processing platform. 
These start-up phases determine a fixed energy cost for every activation of the digital processor.

\begin{table}[]
\caption{System Delay Timing Parameters}	\label{tab:timing}

\centering
\resizebox{1\columnwidth}{!}{
    \begin{tabular}{ | l | p{5cm} | l | }	
    \hline
	Parameter  & Description & Value [$\mu sec$]	\\
	\hline \hline
	$T_{RO}$ & Sensor readout Time & 300 \\
	$T_{ON}$ & Power-on and FLL lock time & 590\\
	$T_{BOOT}$ & Boot-up process time [30 MHz] & 61 \\
	\hline
    	\end{tabular}
}
\end{table}

Fig. \ref{fig:state} depicts the power management strategy implemented within the system at runtime.
When the sensor is in \textit{Idle} mode, 
the other components are put in the lowest power saving mode. In this state, the system still continuously senses the environment thanks to the vision chip and the camera interface operation. Differently from common frame-based vision systems, the digital processor is normally kept in deep-sleep mode and activated depending on the context-activity. According to such event-driven framework, the camera interface switches the sensor to \textit{Active} mode when the amount of sensed data overcomes the pre-defined threshold and then triggers the processor wake-up. Within the interface, the ring-oscillator is only activated within the readout phase, leading to small increment of the average power consumption with respect to the Idle power state. 
After completing the readout, the camera interface issues a wake-up event to the external power manager. This latter requests the power-on of the platform, by triggering an activation signal to both the power-gates of the cluster and the FLL regions.

When powered-on, the PULP power consumption consists of the active power of the described power domains.
After an initial transient period ($T_{ON}$), which also accounts the FLLs activation, the external power manager asserts the fetch-enable signal to start the execution of the application code, stored in the L2 memory. Hence, a time $T_{BOOT}$ is required to set-up the platform before reading data from the camera interface through the SPI interface. During the SPI readout, the FPGA camera interface increases its power consumption by $456\mu W$ due to the peripheral transactions. 
After data transfer and parallel processing, a GPIO signal, acting as EOC, is finally asserted by the PULP platform. Correspondingly,
the external power manager switches back the platform to deep-sleep mode, by turning-off the cluster and FLL regions power-gates.

\section{Always-on monitoring applications}
\label{sec:App}

The visual data coming from the imager are locally analyzed to detect application-specific events of interests. This greatly reduces the output bandwidth of the system with respect to streaming out all the raw sensor data. The flexibility of the processing unit, in terms of software programmability, is crucial to accomplish this task, due to the easy parameter tuning and algorithm adjustments when dealing with different applications. When a detection of an event-of-interest occurs, the smart camera triggers a single-bit alert signal. Depending on the user preference, a control action can take place afterwards, such as counting operations or the activation of a secondary high-resolution RGB camera.

The employed object detection and triggering scheme elaborates the data coming from the contrast-based binary imager. Differently from traditional cameras, the vision chip produces a stream of \textit{(x,y)} addresses, corresponding to the asserted pixels in the binary frame. A lightweight computational framework for processing sensor data and triggering alert signals after the detection of an event-of-interest has been presented in \cite{RusciISC2}. The process is composed of an object detection stage followed by a tracking operation to associate information extracted from successive frames. A trigger signal is generated when a detected and tracked object, described by its center of mass and a surrounding bounding box, matches a predefined condition as, for instance, passing through a virtual loop on the camera plane. 

The first object detection phase operates only on the pixel \textit{(x,y)}-addresses originated by the moving objects on the camera plane. A clustering algorithm is applied to the contrast-based data coming from the sensor to extract the position and the bounding box of moving object. The clustering process starts by the knowledge of the object position on the previous frame and groups the pixels together based on a distance criteria. A first filtering operation is performed on the detected blob based on the number of grouped pixel. Following this, a merge operation between close blobs takes place and a second filtering refines the final list of detected objects.
This approach is inspired by the algorithm proposed in \cite{Litzenberger2006}, which targets object tracking on data coming from event-based asynchronous sensors. Details on a parallel implementation can be found in \cite{Rusci2016}, which also reports the lower computational cost with respect to traditional frame-based object detection approaches.
Following the moving object extraction, detected objects are associated to previously tracked objects. 
The trajectory of every tracked objects is refined by means of Kalman filtering. 

\begin{figure}[]
	\centering
  	\includegraphics[width=0.8\linewidth]{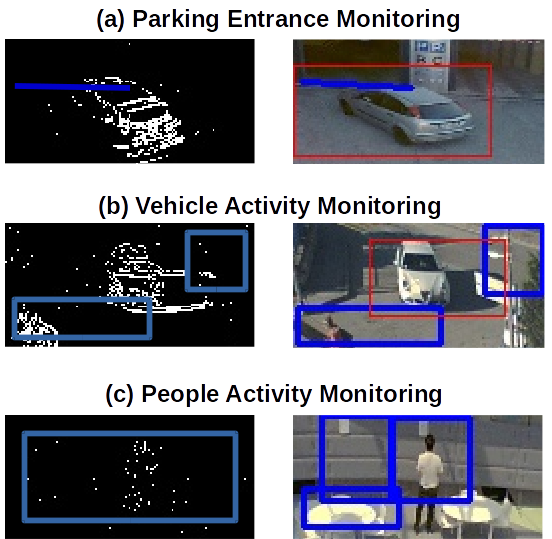}
	\caption{Monitoring scenarios for the considered applications. Frames on the left are captured by our imager (active pixels are drawn in white) and frames on the right come from a commercial RGB  camera. }
	\label{fig:apps_fig}       
\end{figure}

To assess the performance of the presented smart camera system, three environmental monitoring scenarios are considered and listed below. Figure \ref{fig:apps_fig} illustrates some frames captured from each scenario.

\subsection{Parking Entrance Monitoring}
The outdoor entrance of a parking space is monitored to detect cars entering from the left gate. The triggering process is activated when a moving object crosses the virtual gate corresponding to the parking target entrance (Fig. \ref{fig:apps_fig}a). To lower false positive triggering alerts, detected objects with a low number of pixels are filtered out after detection. 

\subsection{Vehicle Street Monitoring}
The smart trigger is employed to monitor vehicles at a crossroad. Here, the triggering condition is associated to any tracked vehicle entering one of the defined virtual loops corresponding to different crossroad directions (Fig. \ref{fig:apps_fig}b). 
Differently from the first scenario, the alert generation is refined to avoid triggering false positive alerts while people crossing the street. More in details, the presence of shadows due to the high illumination makes trigger generation a more tricky task, possibly causing extra alert signal assertions.

\subsection{People Activity Monitoring}
 
The third application deals with people crossing an indoor area and possibly stopping by a landmark or a point of interest (Fig. \ref{fig:apps_fig}c). 
To detect these latter events, an alarm is triggered when any detected and tracked object disappears from the tracking list. Motionless objects are not detected by the sensor because of the internal frame difference operation. Therefore, any tracker associated to moving people is expected to disappear when a person stops. To increase the reliability of the trigger generation, trackers disappearing at the frame borders do not cause any alarm since they are most likely associated to people exiting from the scene. Likewise, limited spatial movements of tracker's center of mass, such as those generated by people standing still, is filtered to avoid generation of false alarms.

\section{Experimental Result}
\label{sec:Eval}

\subsection{Smart Trigger Generation}
The trigger generation process has been evaluated on a dataset of images taken for all the considered applications\footnote{The labelled videos can be found online at: http://e3da.fbk.eu/projects/energy-efficient-smart-vision-systems [accessed on: 2016-09-12]}. The frame rate of the camera is set to 10 frames per second. The evaluation has been initially conducted with nominal environmental conditions (i.e. good lighting).
For every scenario, the camera has been placed in a fixed position to not degrade the motion estimation, which is internally performed on the sensor side through the frame difference operation. During the acquisition sessions, a secondary RGB camera has been used for comparison purpose (Sony Playstation Eye\textsuperscript{TM} camera).

Table \ref{tab:detection} reports the characteristics of the visual dataset for the three monitoring scenarios, in terms of number of frames and occurring events of interest. We manually classify every raised alert signal as True Detection (TD), when an event of interest is correctly detected, False positive (FP), if an unwanted trigger is generated, and False Negative (FN), in case of miss-detection. Based on that, the following metrics are computed for process evaluation:

\begin{equation}
precision = \dfrac{\# TD}{\# TD + \# FP}
\end{equation}

\begin{equation}
recall = \dfrac{\# TD}{\# TD + \# FN}
\end{equation}

where $\#$ refers to the events' cardinality. Note that both the indicators should be equal to 1 for an optimal detection. 

For comparison purpose, we implement a basic object detector on the RGB images, following a common algorithm flow such as \cite{Chen2007}. Every frame is transformed and scaled to match the resolution of our imager. Pixel values are also converted into a grey-scale representation. For any of the considered applications, objects are extracted by computing the connected components after applying a background subtraction, based on a Mixture of Gaussians (for vehicle detection and street monitoring) or frame difference (for people activity monitoring), and morphological filters. 
Blobs with a low number of pixels are filtered out to improve detection accuracy. Similarly to the procedure described in Section \ref{sec:App}, an alert signal is triggered whenever a tracked object matches a predefined condition. Tracking is obtained by associating objects from successive frames based on the centroid's distance and the bounding box's size differences.

\begin{table}[]
\caption{Monitoring application dataset}	\label{tab:detection}

\centering
\resizebox{1\columnwidth}{!}{
    \begin{tabular}{ | l | l | l | }	
    \hline
	Application  & {\#}frames & {\#}events	 \\
	\hline \hline
 	Parking Entrance Monitoring	 &	4000 & 9 \\
 	\hline
	Vehicles Street Monitoring	&	4000 & 63 \\
    \hline
    	People Activity Monitoring 	&	1446 & 19 \\
	\hline 
    	\end{tabular}
}
\end{table}

\begin{figure}[]
	\centering
  	\includegraphics[width=1\linewidth]{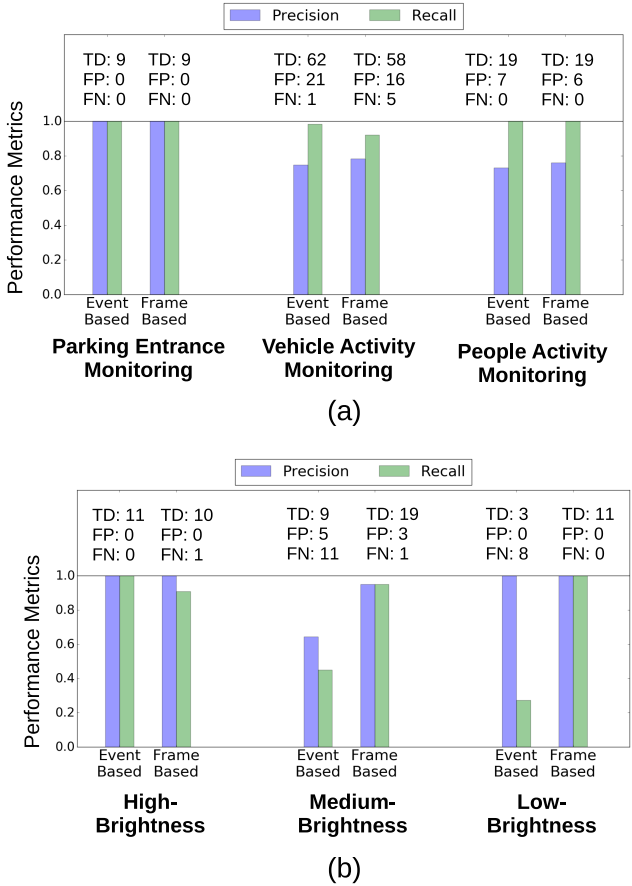}
	\caption{Performance metrics for event-based and frame-based implementations on dataset from (a) monitoring applications and (b) on different lighting exposures}
	\label{fig:metric}       
\end{figure}

Figure \ref{fig:metric}a reports the statistics and the performance metrics computed for the triggering process on the two domains: event-based and frame-based. The first one relates to the triggering process of Section \ref{sec:App} applied on images taken by our vision chip, while the second one refers to the analysis performed on the RGB images. The "Parking Entrance Monitoring" application shows optimal precision and recall on both the domains because of the clear vehicle appearances. On the contrary, the scenario addressed for "Vehicle Street Monitoring" is more challenging. Both the event- and frame- based implementations suffer from a high number of false positive alarms caused by moving people in the scene. Along with it, the event-based approach also presents extra FP alarms due to the fact that a single vehicle can be tracked by multiple blobs, each one attached to different parts of a vehicle. On the other side, the frame-based approach shows a higher precision but lower recall. The increased number of false negatives arises from the loss of trackers due to the merge of close vehicles after background subtraction and morphological filtering.
The "People Activity Monitoring" application presents similar performance metrics for both domain implementations. False Positives are caused mostly by slight movements of people standing still or crossing persons. 

\begin{figure}[b]
	\centering
  	\includegraphics[width=0.8\linewidth]{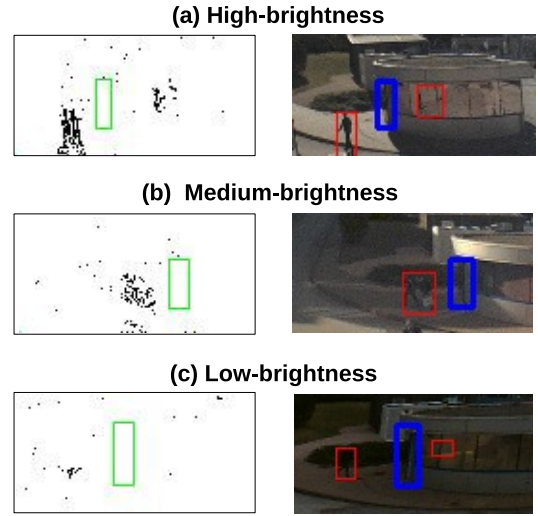}
	\caption{Monitoring scenarios with different light exposure. Frames on the left are captured by our imager (active pixels are drawn in black) and frames on the right come from a commercial RGB  camera.  }
	\label{fig:apps2_fig}       
\end{figure}

To evaluate the robustness of our system to different light exposures, the "People Activity Monitoring" application is tested against an additional video dataset, collected in an outdoor scenario. The considered application involves the monitoring of people entering and exiting from a building gate. A trigger signal is generated whenever a person or a group of compact people enter or exit the virtual loop corresponding to the door. Figure \ref{fig:apps2_fig} shows the targeted scenario with three different light conditions: high-brightness (captured at midday), medium-brightness (early in the morning) and low-brightness (late in the evening). Figure  \ref{fig:metric}b reports the statistic of the trigger process within the above mentioned scenarios, along with a comparison with the same approach in the frame-based domain.
At nominal lighting conditions, the presented system shows a triggering accuracy equal or slightly lower than traditional "frame-based" system. Only for low-light conditions, performance metrics appears degraded with respect to the frame-based approach. This is a consequence of a reduced spatial contrast within the images. Most likely, the object detection processing of Section \ref{sec:App} fails because of the low number of input data. On the contrary, background subtraction of the frame-based approach still performs properly, independently from the environmental lighting. However, this limitation can be mitigated by a refined camera positioning, such as pointing to outdoor places with artificial illumination, and avoiding dark and low-contrast spots. 


\subsection{Power Consumption}
In the following, we report the evaluation of our system in terms of power consumption. To this purpose, measurements has been taken from a device sample in correspondence of the power modes of Section \ref{sec:Prot}. The evaluation is conducted against the application datasets described in Section \ref{sec:App}.

When fully-active, the smart camera system presents a power consumption of 7.62mW, mainly due to the contribution of the PULP platform and the FPGA interface. Individual power measurement values are indicated in Tab. \ref{tab:power}. Thanks to the lightweight application signal processing, the processor is not required to run at the maximum speed all the time. Therefore, as a first optimization step, the digital processor is activated after every frame acquisition and then put in deep-sleep mode when the data processing action completes. Such kind of computational framework is here referred as "Periodic-Polling". In addition to this, the high-speed clock of the FPGA camera interface is enabled only for a short readout period. By applying this power management strategy, the average power consumption of the individual components is extremely reduced. The FPGA camera interface achieves an average power of $77\mu W$, mainly dominated by the consumption associated to the 32kHz clock domain. The processor power in sleep mode is due to the SoC region leakage power of $99\mu W$. In active mode instead, the energy consumption depends on the application computation time, the data transfer operation and the start-up process, which accounts the power-up, the FLL activation and the boot process and represents a fixed cost of every processor wake-up. For frames composed of a minimal amount of pixels, this latter dominates the whole energy cost. 
Table \ref{tab:comp} contains the average power consumptions values on the monitoring application when applying the described power management strategy.
With respect to a fully active system, a power reduction of at least 25x is achieved thanks to the optimized periodic-polling framework.

\begin{figure}[]
	\centering
  	\includegraphics[width=0.6\linewidth]{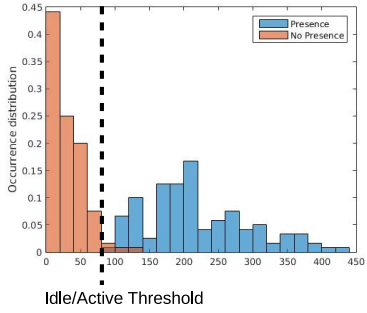}
	\caption{Tuning of the wake-up pixel threshold by observing the distribution of the number of pixel for frames that contains objects (blue class) and not (red class). The class distribution is built by picking 120 uncorrelated samples per class from the "People Activity Monitoring" dataset. }
	\label{fig:thresh}       
\end{figure}

The second optimization step concerns shifting the computational model from the "Periodic-Polling" to the  "Event-Driven" framework, which is enabled by the designed camera interface. The Event-Driven computational model is intended to keep the processor in deep sleep mode until the sensor produces a relevant amount of motion data from the external context.
Consequently, the camera interface transfers data to the processor unit only if the frame contains at least a specific number of meaningful data.
The wake-up threshold value corresponds to the value defined within the filtering stage of the blob detection module. Basically, the system is not awake when any objects will not surely be detected by the post-processing operation because of the low-amount of produced statial-contrast pixels.
The mentioned filtering parameter is manually tuned based on the average amount of pixels describing the objects of interest. Within the "People Activity Monitoring" application, a further analysis is done to investigate a finer tuning. Figure \ref{fig:thresh} shows the probability distribution of the number of pixel in frames belonging to two classes: the red one includes frame with no object presence, while the blue one corresponds to frames with the presence of one or more moving object. Based on that, the threshold is set slightly higher than the filtering value (80 instead of 40), leading to higher power savings because of the reduced number of activations of the digital processing sub-system.

When motion is not detected in the camera field of view, the system exploits the Idle state as described in Section \ref{sec:Prot}. In this state, the vision chip works in Idle and the ring-oscillator of the FPGA camera interface is not enabled. Considering also the processor leakage due the the SoC region consumption, an overall Idle power of $176.88\mu W$ is achieved. Potentially, the average power consumption of the system converges to this bottom limit when no meaningful information is detected within the context. Tab. \ref{tab:comp} highlights the power reduction considering a selective processor activation in correspondence only to the subset of relevant frames (those with a number of asserted pixels higher than the defined threshold). Clearly, a higher reduction is assessed for the application featuring a lower percentage of interesting frames (16\%). In this scenario, a power consumption of $193\mu W$ is achieved, closer to the Idle power lower bound. Instead, within the other application scenarios, a power consumption between $252\mu W$ and $277\mu W$ is reported, coming from a percentage of processor activation slightly higher than 60\%. In these cases, a power saving of about 6\% is achieved if compared with the correspondent periodic-polling computational framework.

\begin{table}[t]
\caption{Power Consumption in the considered applications case-studies}	\label{tab:accuracy}

\centering
\resizebox{1\columnwidth}{!}{
    \begin{tabular}{| l | p{1cm} p{1cm} p{1cm}  |p{1cm} p{1cm} | p{1cm} | }	
    \hline
	Application  & Pixel Threshold & \% relevant frames & App Duty Cycle & Periodic-Polling Power & Event-Driven Power & Reduction \\
	\hline \hline
 	Parking Entrance Monitoring	& 100 &	 16\% & 0.7\% & $226\mu W$ & $193\mu W$ &	-14.6\%\\
 	\hline
	Street Traffic Monitoring & 40 & 60.5\% & 1.1\% & $294\mu W$ & $277\mu W$	& -6\%\\
    \hline
    	People activity Monitoring & 80 & 65.4\%  & 1\%	& $267\mu W$  & $252\mu W$	& -5.6\%\\
	\hline 
    	\end{tabular}
}
\end{table}

\begin{figure}[b]
	\centering
  	\includegraphics[width=1\linewidth]{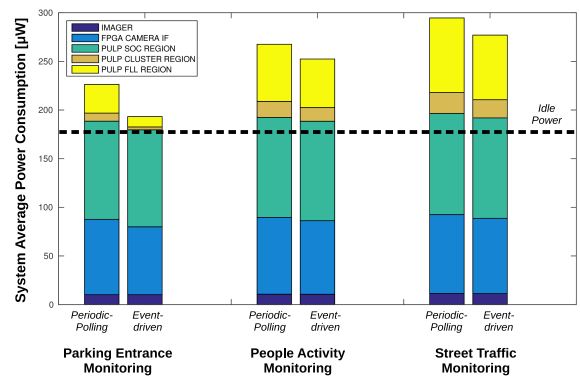}
	\caption{Power consumption breakout of the system }
	\label{fig:energy}       
\end{figure}

\begin{table*}[t]
\caption{Smart Camera Comparison}	
\label{tab:comp}
\centering
\resizebox{1\textwidth}{!}{

	\begin{tabular}{| p{1.8cm} || p{3cm}| p{3cm} | p{3cm} | p{3cm} |p{3cm} | p{3cm} |p{3cm} |}
	\hline
	&  CMUcam5 \cite{CMUCAM} & CITRIC \cite{Casares2011} & MeshEye \cite{Hengstler2007}  &Cyclops \cite{Rahimi2005} & \cite{Carey2013} & \cite{Gasparini2011} & This work \\ \hline \hline
	
	 Image Sensor  & Omnivision OV9715 (1280x800) & Omnivision OV9655 (1280x1024, 640x480) & Agilent ADNS-3060 (30x30), Agilent ADCM-2700 (640x480) & Agilent ADCM-1700 (352x288) & SCAMP \cite{Dudek2006} (128x128 w/ focal plane processing) & Vision Chip \cite{Gottardi2009} (64x128 w/ focal plane processing) & Vision Chip \cite{Gottardi2009} (64x128 w/ focal plane processing)\\ \hline
	 
	 Digital Signal Processing Unit & NXP LPC4330& PXA270 (32-bit) & ARM7TDMI & ATMEL ATmega128L 8-bit & IGLOO FPGA + NXP LPC1769& IGLOO FPGA & IGLOOnano FPGA + PULPv3\\ \hline

	 Visual Processing & Color-based Object Detection & Object Detection and Tracking & Object Detection, Stero matching, object acquisition & Object Detection  & Object detection and counting - Smart trigger & Single people counter & Smart visual Trigger - Object Detection and Tracking\\ \hline
	 Power Consumption & 700mW (typical 140mA @ 5V) & 751mW & 175.9mW \cite{Tavli2012} & 33mW & 5.5 mW & 4.2mW & 277$\mu$W\\ \hline
	 Power Management Strategy& N/A &Image down-sampling and cropping at the hardware level & High-resolution camera acquisition is triggered once and object is detected and stereo matched & CPLD frame-grabber halted by the MCU, as well as the external SRAM & FPGA wakes-up by the MCU after image acquisition & Imager sleep mode - Clock-gating & Camera Interface wakes-up the processor only when relevant information has detected by the imager\\ \hline
	 
	\end{tabular}
}
\end{table*}

Figure \ref{fig:energy} depicts the average power consumption breakdown on the two considered computational frameworks. 
The higher energy-efficiency of the event-driven computational model mostly arises from the power saving exploited within the Cluster region and the FLL region. 
More precisely, a power saving of about 65\% is accounted within the first application in the two mentioned power domain regions. Instead, a 15\% reduction is reported for the other two applications, because of the higher processor activation rate. The saved energy is mostly related to the energy fixed costs regarding the processor wake-up procedure (power-up, FLL activation and boot). 
In addition to this, a FPGA power saving between 5\% and 10\% is accounted, due to less high-speed clock activations and SPI transactions.

Tab. \ref{tab:comp} compares our proposed smart camera system with other existing devices. 
Some of them \cite{CMUCAM,Casares2011,Hengstler2007,Rahimi2005} show a power consumption of tens to hundreds of mW, which is much larger than our targeted consumption due to the power costs of imagers and processing units. 
On the other side, they feature higher-resolution traditional sensors.
The ultra-low-power visual node in \cite{Gasparini2011}, which embeds the same imager of our system, exploits a hardware circuit for visual processing, which limits the application of the node to a single people counter. Unlike this, our system presents a greater flexibility thanks to the fully-programmable processor. With respect to the smart camera presented in \cite{Carey2013}, our system shows a power consumption reduced by $19\mathrm{x}$ in a monitoring application scenario. Differently from it, we employ an optimized event-driven framework that lowers the active power consumption of the system by selectively activating the digital processing unit when an interesting event is detected by the always-on imager.

\section{Discussion and Conclusion}

\label{sec:concl}
In this paper we presented the implementation of a fully programmable ultra-low-power smart camera node. The system is based on a vision chip with analog focal-plane processing that provides the number of asserted pixels when kept in a low-power Idle mode. We reported the implementation on a flash-based FPGA of a dedicated camera interface and the optimized system power management strategy. 

To reach an average power consumption lower than other presented camera systems, the presented design approach exploits an event-driven computational model. Differently from other referenced strategies:
\begin{itemize}
\item the image sensor produces a binary spatial-contrast compressed information with address-event representation,
\item the system power management is driven by the sensor event-activity and it is enabled by an optimized camera interface,
\item the visual processing applies directly on the spatial-contrast sensor data (event-based representation), leading to a computational cost for data processing that depends on the amount of input data \cite{RusciISC2}.
\end{itemize} 
Thanks to all these properties, our approach allows to bring the power consumption as low as the minimum idle power in case of very low context activity. In real application scenarios, our system presents a slightly higher average power consumption between $193\mu W$ and $277\mu W$, which is an order of magnitude lower than other presented smart cameras (Tab. \ref{tab:comp}). Within the considered monitoring applications, we observed that a suitable triggering accuracy is reached by processing the compressed event-based spatial contrast information. For a fair comparison, we compared the triggering accuracy of our approach against a visual processing flow that analyzes RGB or greyscale frame-based signals, which is exploited by state-of-the-art systems. Figure \ref{fig:metric}a reports the metrics comparison, demonstrating that our approach does not lead to performance degradation with respect to other systems. However, when operating in low-light environments, our device is less robust due to the reduction of sensor data produced by the imager (Fig. \ref{fig:metric}b).

This work is a first exploration on an event-driven system-level design approach to sensibly increase energy-efficiency of camera systems. We provided the key ingredients of the proposed strategy and we showed the produced benefits against state-of-the-art systems, claiming that this represents a viable way for visual sensor devices for the IoT ecosystem. The limitation that affects the current prototype will be addressed in future work, motivated by the significant gains achieved thanks to the proposed approach.



%

\section*{Acknowledgment}
This research was funded by the Swiss National Science Foundation under grants 157048 (Transient Computing Systems) and 162524 (MicroLearn: Micropower Deep Learning). The authors would like to thank Massimo Gottardi and Michela Lecca for their insightful discussions.

\ifCLASSOPTIONcaptionsoff
  \newpage
\fi



\bibliographystyle{IEEEtran}
\bibliography{IEEEabrv,jsps}
\end{document}